\documentclass[onecolumn,showpacs,preprintnumbers,amsmathamssymb]{revtex4}
\usepackage{graphicx}
\usepackage{epsfig}
\preprint{V3Si}
\begin{document}

\title{ Multi-band conductivity and multi-gap superconductivity in V$_3$Si from
optical measurements on films at terahertz frequencies}
\author{A. Perucchi$^{1}$, D. Nicoletti$^2$, M. Ortolani$^3$, C. Marini$^2$,
R. Sopracase$^2$, S. Lupi$^{1,2}$, U. Schade$^4$, M. Putti$^5$, I. Pallecchi$^5$,
C. Tarantini$^5$, M. Ferretti$^5$, C. Ferdeghini$^5$, M. Monni$^6$,
F. Bernardini$^6$, S. Massidda$^6$, and P. Dore$^2$}
\affiliation{$^1$Sincrotrone Trieste, Area Science Park, I-34012 Trieste, Italy}
\affiliation{$^2$ CNR-INFM-Coherentia and
Dipartimento di Fisica, Universit\`a di Roma La Sapienza, Piazzale
Aldo Moro 2, I-00185 Rome, Italy}
\affiliation{$^3$CNR-Istituto di Fotonica e Nanotecnologie, Via Cineto Romano 42,
I-00156 Rome, Italy}
\affiliation{$^4$Berliner Elektronenspeicherring-Gesellshaft f\"{u}r
Synchrotronstrahlung m.b.H., Albert-Einstein Strasse 15, D-12489
Berlin, Germany}
\affiliation{$^5$CNR-INFM-LAMIA and Universit\`a di Genova, Via Dodecaneso 33,
I-16146 Genoa, Italy}
\affiliation{$^6$CNR-INFM-SLACS and Dipartimento di Scienze Fisiche,  Universit\`{a} degli
Studi di Cagliari, S.P. Monserrato-Sestu km 0.700, I-09124 Monserrato (Cagliari), Italy}

\date{\today}

\begin{abstract}
The possibility of multi-band conductivity and multi-gap superconductivity is
explored in oriented V$_3$Si thin films by means of reflectance and
transmittance measurements at terahertz frequencies. The temperature dependence of
the transmittance spectra in the normal state gives evidence of two bands
contributing to the film conductivity. This outcome is consistent with
electronic structure calculations performed within density functional theory.
On this basis, we performed a detailed data analysis
and found that all optical data can be consistently
accounted for within a two-band framework, with the presence of two
optical gaps in the superconducting state corresponding to
2$\Delta/kT_c$ values close to 1.8 and 3.8.
\end{abstract}

\maketitle

Large interest has been devoted
\cite{Bouquet02,Mazin03} to multi-band superconductivity
after the discovery of two bands with two distinct superconducting gaps
in MgB$_2$ \cite{Kortus01}.
Very recently, the interest in multi-band superconductivity
has been renewed \cite{Moreo} by the discovery of the Fe-As based
superconductors, where the presence of multiple bands is well
established \cite{Maz08}. It is worth noting that two-gap
superconductivity has been considered theoretically since the fifties
\cite{Suh59}, and experimentally observed in transition
metals \cite{Viel66}, in the A15 compound Nb$_3$Sn \cite{Broc69,Gur04}
and in MgCNi$_3$ \cite{Walt04}.
However, only after the discovery of superconductivity in MgB$_2$, clear cut
evidence of two-gap superconductivity and of its implications was obtained,
since in this compound the multi-band, multi-gap character is emphasized
by the exceptionally low inter-band scattering \cite{Maz02}.

In the case of the A15 V$_3$Si system, a spread in the gap values
corresponding to 2$\Delta/kT_c$ extending from 1.0 to 3.9 has been
reported in the past \cite{Tanner73,Mit86}.
Recently, the electrodynamic response
in the microwave region gave evidence of two gaps \cite{Nef05},
while muon spin rotation measurements were consistent
with a single-gap model \cite{Cal05}. It is also worth noting that
V$_3$Si has been recently treated as a two-band system
\cite{Hauptmann09,Kogan09}, but existing band structure calculations
\cite{Klein78,Borg89} do not provide detailed enough information.

In this contradictory scenario, we investigate in the present work the possible
two-band, two-gap character of V$_3$Si, by performing both
an infrared spectroscopy study and electronic structure calculations
within density functional theory (DFT).

Infrared (IR) spectroscopy is a powerful tool to study the
properties of a conducting system. In the normal state (N-state), the Drude
model for the frequency-dependent complex conductivity
$\tilde{\sigma}_{N}=\sigma_{1N}+i\sigma_{2N}$ can be employed to
describe the optical response of free-charge carriers \cite{Burns}.
As shown in the MgB$_2$ case \cite{Kuz07,Ort08}, it is thus possible
to determine the contributions of different $i$-bands and thus the
corresponding plasma frequencies $\Omega_i$ and scattering rates
$\gamma_i$. In the superconducting state (S-state), on the basis of the BCS
model for the complex conductivity
$\tilde{\sigma}_{S}=\sigma_{1S}+i\sigma_{2S}$ \cite{Tink},
far-IR/terahertz measurements can be of particular importance since
a mark of the superconducting gap $\Delta$ can be observed at
$\hbar\omega \sim 2\Delta$ (optical gap) for an isotropic $s$-wave
BCS superconductor. In particular, a maximum at the optical gap is
expected either in the ratio $R_S/R_N$ (for a bulk sample) or in
the ratio $\mathcal{T}_S/\mathcal{T}_N$ (for a thin film),
where $R_S$ ($\mathcal{T}_S$) and $R_N$ ($\mathcal{T}_N$) are the
frequency-dependent reflectances (transmittances) in the
S- and N- state, respectively \cite{Tink}.
In the MgB$_2$ case \cite{Kuz07},
evidence of the two gaps was detected in the $R_S/R_N$ spectrum
of an ultra-clean film \cite{Ort08} by assuming a parallel sum of the
conductivity of two independent bands and by using the model
$\tilde{\sigma}_{S}$ introduced by Zimmermann \emph{et al.} \cite{Zim91}
which generalizes the BCS one to arbitrary $T$ and $\gamma$ values
(Zimmermann model).

We performed IR measurements on high-quality V$_3$Si textured films.
Details on the film growth by pulsed laser deposition and on their
properties are reported elsewhere \cite{Ferd1}.
We studied two films grown on LaAlO$_3$ (LAO) (001) 0.5 mm thick substrates,
which exhibit preferential (210) orientation along the out-of-plane direction.
The first film of thickness $d$=180 nm (film d180) has good
transport properties (resistivity at 300 K close to 200 $\mu\Omega $cm,
residual resistivity ratio $RRR$=8) and $T_c$=16.1 K; the second
film, 33 nm thick (film d033), has worst transport properties
($RRR$=4.5) and a slightly lower $T_c$ value ($T_c$=15.3 K).

We first performed measurements of the IR reflectance $R(\omega)$ of
film d180 as a function of temperature (not shown). The $R(\omega)$
spectrum at 300 K is in very good agreement with the result of
previous measurements \cite{Borg89}. We attempted an analysis of the
measured spectra by including in the model $\tilde{\sigma}_{N}$ a
number of Lorentz contributions representing optically active
inter-band transition. However, unambiguous results were not
obtained since at any temperature a very broad Drude contribution
\cite{Borg89} is partially overwhelmed by strong interband
transitions occurring at rather low frequencies because of the
complex electronic structure of V$_3$Si \cite{Klein78, Borg89}.

\begin{center}
\begin{figure}
\psfig{figure=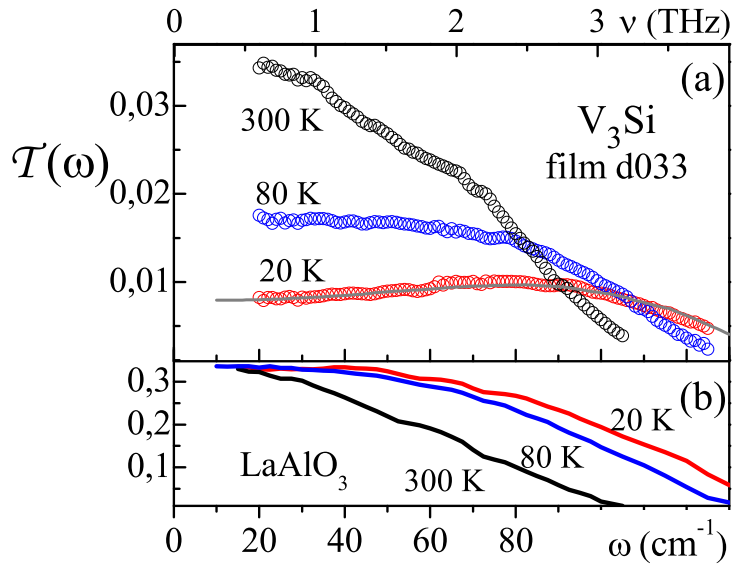,width=10cm} \caption{(Color online) (a)
Transmittance spectra $\mathcal{T}(\omega)$ of film d033 at selected
temperatures in the THz region. The 20 K spectrum is compared with
the two-band (2-b) best fit. (b) $\mathcal{T}(\omega)$ spectra of a
LaAlO$_3$ substrate (0.5 mm thick) at the same temperatures
\cite{Dore94}.} \label{Transmittance_N}
\end{figure}
\end{center}

We then measured the transmittance $\mathcal{T}(\omega)$ spectrum of
film d033 by focusing the IR beam on the film or on a hole for
reference. Measurements were performed between 300 and 20 K in the
terahertz (THz) region (here defined by photon energies below 16
meV, frequency $\omega<$ 130 cm$^{-1}$ or $\nu<4$ THz), where the
LAO substrate is partially transparent. $\mathcal{T}(\omega)$
spectra of film d033 are reported in Fig. \ref{Transmittance_N}a at
selected temperatures, those of LAO \cite{Dore94} at the same
temperatures in Fig.\ref{Transmittance_N}b for comparison. The
spectral shape of the 300 K $\mathcal{T}(\omega)$ spectrum is
qualitatively similar to that of LAO as it monotonically increases
with decreasing frequency, while the transmitted intensity is
strongly reduced by the above discussed broad Drude contribution in
the film conductivity. On the contrary, the spectral shape of the 20
K $\mathcal{T}(\omega)$ shows a broad maximum and then decreases
with decreasing frequency. This result can be qualitatively
explained by the presence of a second sharp Drude contribution,
which narrows with decreasing temperature thus reducing the
transmitted intensity at very low frequencies.

\begin{center}
\begin{figure}
\psfig{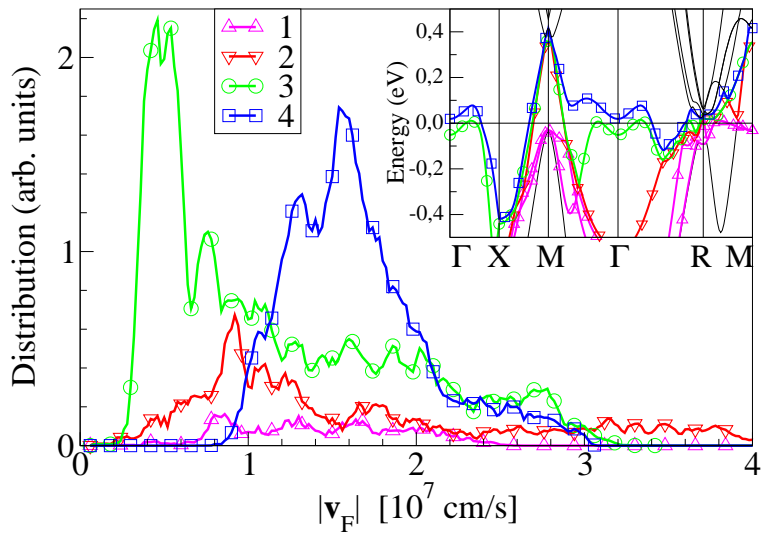} \caption{(Color online)
Distribution of moduli of the Fermi velocities for the bands
crossing $E_F$. Inset: Energy bands along the symmetry lines of the
Brillouin zone; the different symbols identify the bands as in the
inset. \label{fermi_v}}
\end{figure}
\end{center}

In order to better understand this result, we carried out electronic
structure calculations within DFT on V$_3$Si. We calculated the
Fermi velocities starting from the energy bands shown in the inset
of Fig.\ref{fermi_v}. We used the experimental lattice constants $a$
and $c$ \cite{Chaddah83}, and calculated the dimerization of V
chains along $x$ and $y$ directions to be $0.0025 a$. The Fermi
level $E_F$ of V$_3$Si is crossed by four bands (two bands with
similar character and small density of states are grouped together)
with mostly V $3d$ character; with the chain direction along $z$,
orbitals $d_{x^2-y^2}$ and $d_{z^2}$ on one side and $d_{xz}$ and
$d_{yz}$ on the other produce narrow and wide bands, respectively.

In Fig. \ref{fermi_v} we plot the distribution of values of the moduli
of Fermi velocities \cite{nota} for the
bands crossing $E_F$; the distributions are
normalized to the densities of states of the same bands, so that the
average squared velocity times the integral of the distribution
provides, apart from numerical factors, the contribution to the
square of the total plasma frequency $\Omega_{tot} =3.34$ eV.
While it is difficult to assign a definite orbital character to each band
throughout the whole Brillouin zone, we can see that the band separation
naturally provides different average velocities, showing an overall bimodal
distribution given by the bands n. 3 and 4, while the remaining bands
provide a smaller contribution to plasma frequencies due to their
smaller density of states.
The bimodal distribution resulting from our calculations indicates that the
V$_3$Si conductivity in the N-state
can be safely described in terms of two bands
characterized by different $\Omega$ values (two-band, 2-b model).
It remains to be verified whether the presence of the two bands reflects in
the opening of two distinct superconducting gaps, as in the MgB$_2$ case.

\begin{center}
\begin{figure}
\psfig{figure=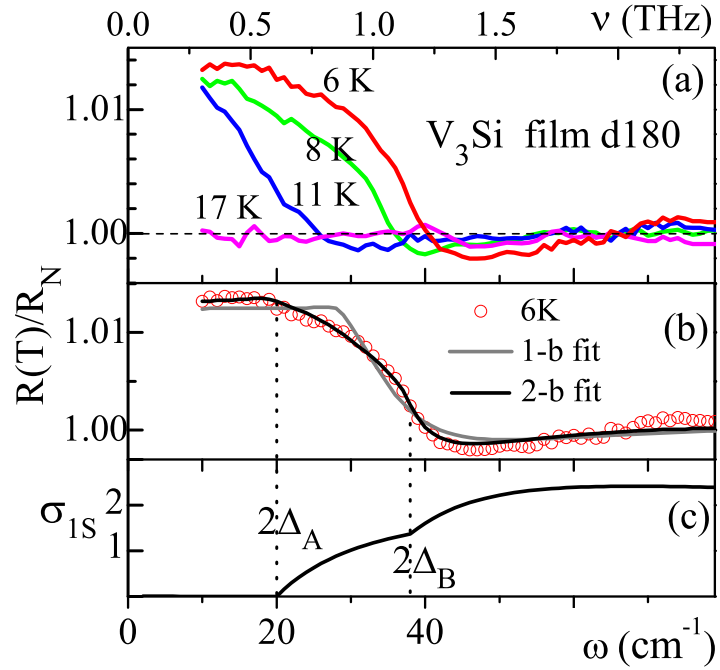,width=10cm} \caption{(Color online) (a)
$R(T)/R_N$ spectra of film d180 at selected temperatures in the Thz
region. (b) $R_S/R_N$ (with $R_S=R(6 K))$ spectrum compared with the
two-band (2-b) and one-band (1-b) best fits. In the 1-b case,
best-fit parameters are $\Omega=2.35$ eV, $\gamma=125$ cm$^{-1}$,
and $\Delta=14$ cm$^{-1}$. (c) $\sigma_{1S}$ (in units 10$^4$
$\Omega^{-1}$cm$^{-1}$) of V$_3$Si from 2-b model.}
 \label{Reflectance_SC}
\end{figure}
\end{center}

\begin{figure}
\begin{center}
\psfig{figure=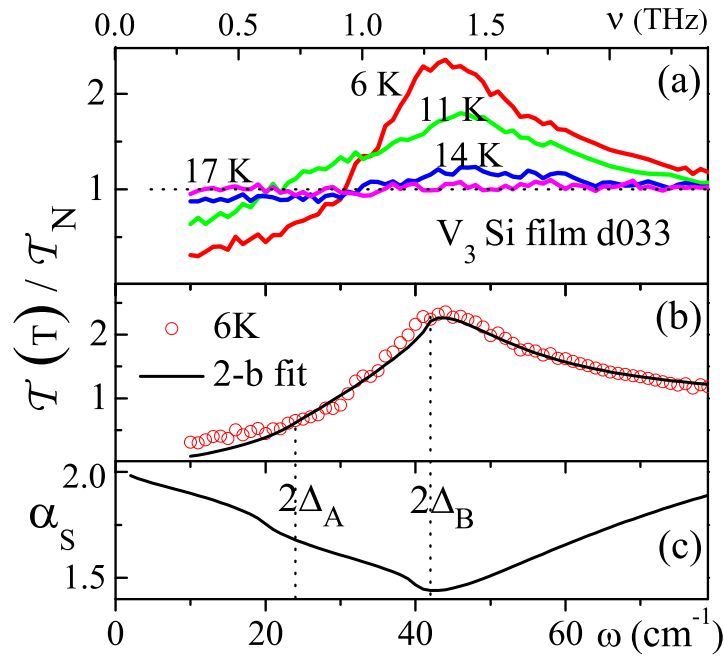,width=10cm} \caption{(Color online) (a)
$\mathcal{T}(T)/\mathcal{T}_N$ spectra (with
$\mathcal{T}_N$=$\mathcal{T}$(20 K)) of film d033 at selected
temperatures in the THz region. (b) $\mathcal{T}_S/\mathcal{T}_N$
spectrum compared with the two-band (2-b) best fit ($\Omega_A$ =
3.35 eV, $\gamma_A$= 1000 cm$^{-1}$, $\Delta_A$= 12 cm$^{-1}$,
$\Omega_B$ = 1.24 eV, $\gamma_B$= 75 cm$^{-1}$, $\Delta_B$= 21
cm$^{-1}$). (c) absorption coefficient $\alpha_S$ (in units 10$^4$
cm$^{-1}$) of V$_3$Si from 2-b model.} \label{Transmittance_SC}
\end{center}
\end{figure}

To this aim we measured at $T\leq$ 20 K the $R_S(T)/R_N$ (where
$R_N=R$(20 K)) spectra of the d180 film (see
Fig.\ref{Reflectance_SC}a) and the $\mathcal{T}_S(T)/\mathcal{T}_N$
(where $\mathcal{T}_N=\mathcal{T}$(20 K)) spectra of the d033 film
(see Fig.\ref{Transmittance_SC}a). These measurements were made by
cycling the temperature in the 6-20 K range, without collecting
reference spectra. In this way one avoids any variation in the
sample position and orientation, which may yield frequency-dependent
systematic errors in $R(\omega)$ and $\mathcal{T}(\omega)$ \cite{Ort06}.
All measurements in the THz region were made by employing synchrotron
radiation at the infrared beamline SISSI \cite{SISSI} at the
synchrotron Elettra (Trieste, Italy) and at the BESSY storage ring
(Berlin, Germany), where coherent synchrotron radiation is
available \cite{Scha01}.
We remark that a high-flux synchrotron source allows an
high accuracy in the detection of small effects in the $R(T)/R_N$
spectra, and overcomes the problem of the very low intensity
transmitted by the film+substrate system in the case of transmission
measurements on a conducting film. We also note that
$\mathcal{T}_S/\mathcal{T}_N$ (where $\mathcal{T}_S=\mathcal{T}$(6
K)) is affected by the superconducting transition much more
than $R_S/R_N$ (where $R_S=R$(6 K)) since the transmitted intensity
is dominated by the effect of the absorptive processes in the film,
which significantly changes below $T_c$ \cite{Tink,Will90}.

We first notice that the $R(T)/R_N$ spectrum increases on decreasing
temperature (see Fig. \ref{Reflectance_SC}a), until $R_S/R_N$
reaches a maximum and becomes nearly constant
below 20 cm$^{-1}$. This indicates the presence of a superconducting
gap $\Delta$ close to 10 cm$^{-1}$. Indeed, for
$\omega\rightarrow0$, the reflectance $R_N$ of a conducting system
tends to 1 for a bulk system, to a slightly lower value
for a thin film. Therefore, since $R_S$ approaches 1 at
$\omega=2\Delta$, $R_S/R_N$ exhibits a maximum around $2\Delta$ in
the case of a bulk sample, remains nearly constant below $2\Delta$
in the film case. As to the $\mathcal{T}(T)/\mathcal{T}_N$ data (see
Fig. \ref{Transmittance_SC}a), a maximum develops on decreasing $T$ until the
$\mathcal{T}_S/\mathcal{T}_N$ exhibits a well defined peak around 40
cm$^{-1}$, which indicates the presence of a superconducting gap
$\Delta$ around 20 cm$^{-1}$.

For a detailed analysis of the measured spectra based on the 2-b
model, the conductivity of V$_3$Si can be described by the sum of
two Drude terms in the N-state ($\tilde{\sigma}_{N}$, with
parameters $\Omega_{i}, \gamma_{i}$, $i=A,B$), and by the sum of two
Zimmermann terms in the S-state, as discussed above
($\tilde{\sigma}_{S}$, with parameters $T$,
$\Omega_{i},\gamma_{i},\Delta_{i}$, $i=A,B$). From $\tilde{\sigma}$,
by using standard relations \cite{Burns}, it is possible to compute
the model refractive index $\tilde{n}=n+ik$ of V$_3$Si in both the
N- and S-state. The transmittance and reflectance spectra of the
film+substrate system in both states can then be evaluated by means
of an exact procedure \cite{Berb93} which requires, besides
thickness, numerical values of $n$ and $k$ for both film and
substrate. For LAO, $n$ and $k$ values in the THz region were
obtained from previous measurements of transmittance and reflectance
of a LAO substrate \cite{Dore94} through a numerical, model
independent procedure \cite{Cuns92}.

We first analyzed the $R_S/R_N$ spectrum, by performing a 6
parameter fit ($\Omega_i$,$\gamma_i$, $\Delta_i$, $i=A,B$). The
best-fit curve does well describe experimental data, as shown in
Fig.\ref{Reflectance_SC}b. By keeping into account the uncertainties
resulting from the fitting procedure, we can safely pose
$\Omega_A=3.7\pm0.1$ eV,\ $\Omega_B=1.00\pm0.05$ eV,
$\gamma_A=800\pm40$ cm$^{-1}$, $\gamma_B= 45\pm4$ cm$^{-1}$,
$\Delta_A=10 \pm 1$ cm$^{-1}$, and $\Delta_B=19\pm2$ cm$^{-1}$. We
verified that unsatisfactory results are obtained by considering one
band only (one-band, 1-b model). Note that, in the 2-b approach, the
A-band contribution has a crucial effect, as it gives a
vanishing $\sigma_{1S}$ only below 2$\Delta_A$ (see
Fig.\ref{Reflectance_SC}c). Since, in general, $R_S$ approaches 1
when $\sigma_{1S}$ vanishes, this explains why only the $\Delta_A$
gap is well evident in the $R_S/R_N$ spectrum.

In order to verify the compatibility of this fit with first
principle results, we calculated plasma frequencies from our Fermi
velocities. In particular, we integrated over the
Brillouin zone separating the low and high ${\bf v}_F$ regions (setting an
arbitrary but plausible cutoff at 10$^7$ cm/s; see Fig.\ref{fermi_v}). While
this procedure is only qualitative, it may be justified by
the presence of clearly separable ranges of ${\bf v}_F$, in turn
deriving from different orbital natures. We obtain $\Omega_A$=3.21
eV and $\Omega_B$= 0.81 eV. In comparing these results with those of
the fitting procedure, it is worth noting that the best-fit values of
$\Omega_A$ and $\Omega_B$, and thus of $\Omega_{tot}$, can be
overestimated because the interband transitions discussed above may
give non negligible contributions to $\tilde{\sigma}_{N}$ even
at very low frequencies. This effect is minimized in the best-fit
$\Omega_A$/$\Omega_B$ ratio (3.7$\pm$0.3), which results to be in a
remarkably good agreement with the computed one (3.96).

In analysing the transmittance spectra of film d033, we verified
that only the 2-b model well describes $\mathcal{T}(\omega)$ at all
temperatures, but the resulting $\Omega_i$ and $\gamma_i$ values are
not unambiguously determined. We thus used a procedure which
simultaneously fits both $\mathcal{T}_N$ and
$\mathcal{T}_S/\mathcal{T}_N$, thus imposing important constraints
in the fitting procedure. Good fits of both $\mathcal{T}_N$ (see
Fig.\ref{Transmittance_N}b) and $\mathcal{T}_S/\mathcal{T}_N$ (see
Fig.\ref{Transmittance_SC}b) were obtained, with $\Omega_i$ values
slightly lower, $\gamma_i$ values slightly higher than those found
in fitting the $R_S/R_N$ spectrum. This can be simply explained by
the worse conducting properties of film d033 with respect to film
d180. As to the gap values, nearly equivalent fits are obtained for
$\Delta_A$ ranging from 11 to 16 cm$^{-1}$, while $\Delta_B$= 21.0
$\pm$0.5 cm$^{-1}$ is well determined since it corresponds to the
peak in $\mathcal{T}_S/\mathcal{T}_N$. The transmission measurement,
dominated by the absorption process, thus permits to unambiguously
establish the $\Delta_B$ value which was more poorly defined in the
$R_S/R_N$ measurement. The higher sensitivity of
$\mathcal{T}_S/\mathcal{T}_N$ to the $\Delta_B$ gap is a consequence
of the frequency dependence of the absorption coefficient $\alpha_S$
of V$_3$Si in the S-state. Indeed, $\alpha_S$, as evaluated with
standard relations \cite{Burns} through the 2-b model, only exhibits
a well defined minimum around 2$\Delta_B$ (see
Fig.\ref{Transmittance_SC}c).

In summary, we addressed the debated problem of multi-band,
multi-gap nature of V$_3$Si by means of reflectance and
transmittance measurements in the THz region on high quality
oriented films. Experimental results indicate the presence of two
bands contributing to the V$_3$Si conductivity in the normal state,
and of two optical gaps in the superconducting state. Electronic
structure calculations within density functional theory showed that
the distribution of the modulus of the Fermi velocity exhibits a
clear bimodal character, which indicates that the V$_3$Si
conductivity in the normal state can be safely described in terms of
two bands, characterized by different plasma frequencies. On this
basis, we performed a detailed data analysis and found that all
optical data can be consistently accounted for within a two-band
framework, with the presence of two optical gaps corresponding to
2$\Delta/kT_c$ values close to 1.8 and 3.8.

The authors acknowledge M. Prasciolu for preparing the Au reference
surface used in the absolute reflectance measurements.

\end{document}